\documentclass[aps,prd,twocolumn,nofootinbib]{revtex4}
\usepackage{amsmath}
\usepackage{epsfig}
\usepackage{subfigure}
\usepackage{float}
\usepackage{longtable}

\newcommand{\beq}[1]{
\begin{equation}\label{#1}}
\newcommand{\eeq}{\end{equation}}
\newcommand{\bea}[1]{
\begin{eqnarray}\label{#1}}
\newcommand{\eea}{\end{eqnarray}}
\newcommand{\out}{\raise-3pt\hbox{\scriptsize    out}}

\begin{document}

\title{Hard exclusive processes with photons}

\author
{B.~Pire$^{a}$,\,
 L.~Szymanowski$^{b,c}$
}

\preprint{CPHT--  hep-ph/yymmnnn}

\affiliation{
$^{a}$Centre de Physique Th\'eorique, \'Ecole Polytechnique, CNRS, 
91128 Palaiseau, France
\\$^{b}$Interactions Fondamentales en Physique et Astrophysique, Universit\'e de Li\`ege,
17 All\'ee du 6 Ao\^ut, B\^atiment B5a, B-4000 Li\`ege-1, Belgium\\
$^{c}$Soltan Institute  for   Nuclear  Studies,  Warsaw,   Poland 
}

\begin{abstract}
Virtual photons have proven to be very efficient probes of the hadronic structure, mostly
through deep inelastic scattering and related processes. The advent of high luminosity lepton beams 
has allowed to enlarge the studied processes to hard exclusive reactions, such as deeply virtual 
Compton scattering and the electroproduction of mesons. We discuss theoretical progress which has lately been 
quite remarkable in this domain and  first much encouraging experimental data. 
\end{abstract}
\maketitle

\section{New factorizations}
A considerable amount of theoretical and experimental work is
currently being devoted to the study of exclusive reactions with a large scale often provided by the 
virtuality of an incoming photon \cite{url}, \cite{review}. Much of it is born out from the
introduction of generalized parton
distributions \cite{GPD}, which are defined as Fourier transforms of
matrix elements of non-local operators on the light-cone between different hadron states, such as (here and below, 
we omit the color indices and the Wilson lines which are required by  the QCD-gauge
invariance):
\begin{equation}
 \langle N'(p',s') |\bar\psi(\lambda
n)(\gamma \cdot n)
\psi(0) | N(p,s) \rangle \;.
\end{equation}
Their crossed version,the generalized distribution amplitudes (GDA) \cite{GDA} 
describe the exclusive hadronization of a
$q \bar q$ or $g g$ pair in a pair of hadrons, a pair of $\pi $ mesons for
instance. These GDAs are
defined in the quark-antiquark case, as Fourier transforms of matrix elements such as
\begin{equation}
\langle \pi(p') \pi(p) |\bar\psi(\lambda n)(\gamma \cdot n)
\psi(0) | 0) \rangle
\label{def}
\end{equation}
being a nontrivial function of
 $W^2$, the squared energy of the $\pi \pi$ system,
$W^2=(p+p')^2$.
They are the non perturbative parts which enter in a factorized way in the
amplitudes of the light cone dominated processes
$$
\gamma^* \gamma \to \pi \pi\;,
$$
which may be measured  in electron positron colliders of high
luminosity and
$$
\gamma^* N \to \pi \pi N'\;,
$$
which is currently explored at HERMES, Compass and JLab.
The GDAs take into account
the final state interactions of the meson pair. Their phases are related to the
 phases of $\pi \pi $ scattering amplitudes, and thus contain information on
the resonances which may decay in this channel.

The measurements of GPDs and GDAs are expected to yield important
contributions to our understanding of how quarks and gluons assemble
themselves into hadrons.

A new    scaling  regime has also been studied \cite{TDA} in   the case of  nucleon   -
anti-nucleon  annihilation into a  lepton pair of virtuality Q and  a meson,  
$$\bar p N \to \gamma ^* \pi $$
 in the forward (or backward) kinematics, where
$|t|<< Q^2 \sim s$  (or $|u|<< Q^2 \sim s$) .
 In this kinematical region  the leading twist amplitude factorizes into 
an antiproton distribution amplitude, a
short-distance matrix element related to 
nucleon form factor and a long-distance dominated 
transition distribution amplitudes (TDA) which describe the nucleon to
meson transition. This TDA is defined from the matrix elements of non-local operator built from three quark fields
on the light-cone
\begin{equation}
   \langle \pi|\, {\psi}^{\alpha}(z_{1}n)\, 
{\psi}^{\beta}(z_{2}n)\, 
   {\psi}^{\gamma}(z_{3}n)\,|p \rangle  .
 \end{equation}  
The same TDAs appear in the description of exclusive channels associated with charmonium 
production 
$$\bar p N \to \;J/\psi \;\;\pi\;, $$
and in the backward electroproduction of a meson
$$\gamma ^* N \to N' \pi\;,$$  
which is being studied at Jlab and Hermes.

The $\bar p N \to \gamma^* \pi $ 
amplitude at small momentum transfer is then proportional to the 
TDAs $T(x_{i}, \xi, t)$, 
where $x_i$ (i=1,2,3) denote the light cone
momentum fractions carried by participant quarks, the sum of which is  related to the
skewedness parameter $\xi = \frac{x_1 + x_2 + x_3}{2}$ connected  to the 
external variables $Q^2$ and the squared center of mass energy $W^2$ by
\begin{equation}
\xi \approx \frac{Q^2}{2W^2-Q^2}
\end{equation}
in the generalized Bjorken limit. The scattering amplitude reads schematically
\begin{equation}
 \int dx dy \phi(y_i,Q^2)
T_{H}(x_i, y_{i}, Q^2) T(x_{i}, \xi, t, Q^2)\;,
\label{amp}
\end{equation}
where $\phi(y_i,Q^2)$ is the antiproton distribution amplitude and
$T_{H}$ the
hard scattering amplitude, calculated in the colinear approximation.

\section{Hadron tomography }
A very interesting peculiar feature of these non-forward matrix elements is the fact that
 they open an access to the transverse localization of quarks and gluons in hadrons. The case of GPDs 
   has  been the subject of intense investigation \cite{Femto1}. Indeed, apart from the longitudinal 
momentum fraction variables, GPDs also depend on the momentum transfer $t = \Delta^2$ 
between the initial and final hadrons ($\Delta = p_{\pi}-p_{p}$). Fourier transform of GPDs with respect to the transverse of $\Delta$, $\Delta_T$, leads to the definition of an impact parameter representation of the GPD
\begin{equation}
 \int \frac {d^2\vec \Delta_{T}}{4\pi^2} e^{-i\vec \Delta_{T}.\vec b_{T}} 
H(x, \xi, t) \;
\end{equation}
which contains information on the transverse position of quarks and
gluons in the hadron. Real-space images of the target in the transverse plane can thus be obtained,
with a spatial resolution of the order $1/Q$, determined by the virtuality of the incoming photon.
Let us recall that the longitudinal size of the target is squeezed by the Lorentz contraction factor so
that the longitudinal position of quarks remains unresolved.

Similar arguments can be developed
for the $W^2$ dependence of the GDAs and the $t-$ dependence of TDAs \cite{Femto2}. A Fourier
transform of their $2-$dimensional transverse momentum behavior   leads to a
transverse impact coordinate $\vec b_{T}$ representation. Let us stress the complementarity of the 
information encoded in the impact parameter representations of GPDs and TDAs. In the ERBL region
(where the quark and antiquark  emerge from the hadronic blob), the proton
 GPDs map out the transverse localization of $\bar q q$ pairs of typical transverse size of order $1/Q$.
 On the other hand, in the ERBL region of the proton to pion transition (defined there by the fact that  three quarks emerge from the hadronic blob)
 TDAs map out the localization of a $q q q$ triplet of typical transverse size of 
 order $1/Q$ in a proton, the remnants of which emerge as a  $\pi$ meson.

\section{Where is asymptopia ?}
Data presented at this conference \cite{Data} and elsewhere \cite{Data2} show that nature is kind with us 
in the sense that early scaling seems
to be the dominant feature of dVCS experiments. This is particularly exemplified through the harmonic analysis of
the azymuthal dependence, which is a specific feature of the interference of the dVCS amplitude with the Bethe-Heitler 
process \cite{DGPR}. The same trend appears also in the analysis of the $Q^2$ dependence of the reactions described 
with generalized distribution amplitudes \cite{LEP} although more precise data are obviously required, which should 
be possible in high luminosity B factories.

The analysis of meson electroproduction data in terms of leading twist contributions is more dubious. 
Next to leading order calculations are now performed \cite{NLO}
and different optimization procedures are being discussed \cite{BLM}. Various strategies for extracting the 
generalized parton distributions from the data have been put forward \cite{extract}  but more 
work is required before one can draw any firm conclusion. 

\section{Exotic hadrons}
Hard production of hybrid mesons which go beyond  the quark model description of hadrons but are
predicted to exist by most QCD inspired models has seemed to be a challenge, since 
such {\em exotic} particles were generally believed to have a vanishing leading
twist distribution amplitude (DA). We demonstrated recently \cite{Hybrid} that the non local nature 
of the quark correlators defining a DA was in fact allowing 
both: non-exotic $J^{PC}=1^{--}$ values as in the case of $\rho$-meson, as well as  exotic values $J^{PC}=1^{-+}$ 
corresponding to a  hybrid meson, i.e. one which involves quark and gluon constituents.
 Moreover, a relation between the energy-momentum tensor and a moment of this DA
allowed us to derive the magnitude of the leading twist DA of a $J^{PC}= 1^{-+}$ exotic
vector meson. Electroproduction cross sections estimates then turn out to be not small in comparison with those of 
usual mesons. We thus believe that precise data at JLab or Compass should be able to reveal the properties of these exotic 
mesons. 

On the other hand, tetraquarks states which have been much discussed recently with respect to heavy resonances
may have quantum numbers incompatible with leading twist distribution amplitudes. This is the case for 
isotensor states (since $\bar q q $ or $g g$ states can only be isosinglet or isotriplet) which then 
may be revealed \cite{Isotensor} by the presence of a twist 4 component in the amplitude of the  
 $\gamma^* \gamma \to \rho \rho$ LEP data with the $Q^2$ behaviour of the ratio 
 $$\frac{d\sigma (\gamma^* \gamma \to \rho^0 \rho^0)}{d\sigma (\gamma^* \gamma \to \rho^+ \rho^-)}$$
 as a specific signature . Confirmation
of such an analysis requires to detect charged isopartners of these states.

\section{QED gauge invariance and factorization}
The gauge invariance property of QED leads to consider factorization properties
with some care when real photons appear in the final state of the reaction.
This is particularly obvious when one tries to define \cite{FPS} generalized parton distributions in a real 
photon. Considering indeed the six diagrams which contribute at Born order to the deeply virtual 
Compton scattering amplitude on a photon target
$$\gamma^* \gamma \to \gamma \gamma$$
one sees that they all contribute at the same order and that they are all needed to ensure
ultraviolet convergence of the amplitude. This superficially looks inconsistent with the factorization
property and the dominance of handbag diagrams. A proper understanding of the meaning of 
factorization in that {\em anomalous} case requires to consider properly renormalization 
scale dependence and  a judicious factorization scale choice.

The factorization of the virtual Compton scattering amplitude on a nucleon requires also some 
care to properly insure  QED gauge invariance \cite{Gauge}. Indeed, the twist expansion of the amplitude is not
trivial for non strictly forward kinematics in the sense that a QED gauge invariant expression requires 
to include the contributions from both leading twist quark correlators and twist 3 quark-gluon correlators :
\begin{equation}
 \langle N'(p',s') |\bar\psi(\lambda
n)(\gamma \cdot n)\;g A_{\rho}\;
\psi(0) | N(p,s) \rangle \;.
\end{equation}
Although the contribution to the scattering amplitude of these new generalized parton 
distributions is proportionnal to the transverse momentum of the outgoing hadron, a consistent 
calculation cannot arbitrarily put them to zero. Thus, the apparent simplicity of the dVCS process
as compared to the meson electroproduction case is blurred by the gauge nature of the photon. 

The picture looks even worse when we go to the case of backward kinematics, where hadron
to photon TDAs describe the factorized soft part of the scattering amplitude. The exemplary case of 
dVCS on a meson is worth a detailed study. If one insists on a gauge invariant description of the process,
one should separate a "structure dependent" contribution from an "inner bremsstrahlung" process, as in
 the much studied case of the radiative weak decay of the $\pi^+$ meson ($\pi^+ \to e^+ \nu \gamma$).
 The structure dependent term may be factorized with the help of $\pi^+ \to \gamma$ TDAs, but the 
 "inner bremsstrahlung" contributions where the real photon is emitted  from the initial or final meson 
 must be calculated in a consistent hadronic model (eventually including seagull terms) without separating 
 a short distance subprocess from long distance matrix elements \cite{Bro}.
\section{ Acknowledgments}
We acknowledge many discussions and common research with I. Anikin, J.P. Lansberg, O. Teryaev and
S. Wallon. This work is supported by the  Joint Research
Activity "Generalised Parton Distributions" of the european I3 program
Hadronic Physics, contract RII3-CT-2004-506078, the ECO-NET program 12584QK and the Polish Grant 
1 P03B 028 28. 
L.Sz. is a Visiting Fellow of the Fonds National pour la Recherche Scientifique (Belgium).


\end{document}